\documentclass[amsmath,twocolumn]{revtex4}
\bibpunct{[}{]}{;}{a}{,}{,}%

\usepackage{graphicx}

\begin{document}

\title{Full-scale simulation study of a generalized Zakharov model for
the generation of topside ionospheric turbulence}

\author{B. Eliasson}
\affiliation{Department of Physics, Ume{\aa} University, SE-901 87
Ume{\aa}, Sweden} \affiliation{Theoretische Physik IV,
Ruhr--Universit\"at Bochum, D-44780 Bochum, Germany}

\begin{abstract}
We present a full-scale simulation study of a generalized Zakharov
model for the generation of the topside electrostatic turbulence due
to the parametric instability during ionospheric heating experiments
near the F region peak. The nonlinear tunneling of electromagnetic
waves through the ionospheric layer is attributed to multiple-stage
parametric decay and mode-conversion processes. At the bottomside of
the F region, electrostatic turbulence excited by the parametric
instability results in the conversion of the ordinary (O mode) wave
into a large amplitude extraordinary (Z mode) wave tunneling through
the F peak. At the topside interaction region, the Z mode undergoes
parametric decay cascade process that results in the generation of
the topside electrostatic turbulence and then conversion process
yielding O waves that escape the plasma. This study may explain the
observed topside ionospheric turbulence during ground based
ionospheric heating experiments.
\end{abstract}

\maketitle
%
%

%


%
%

\section{Introduction}

There have been several radar observations of enhanced plasma and
ion lines at the topside ionosphere during ground
based heating experiments in which the transmitting frequency
has been kept somewhat lower than the critical plasma frequency of the
F2 layer maximum. A proposed explanation for the observed topside plasma and ion lines is
a nonlinear decay of large amplitude Z mode waves, which have been
mode converted from the O mode heater wave at the bottomside of the
ionosphere, into Langmuir and ion acoustic (IA) waves. 
The observations have attracted much interest
since both the ordinary (O) and fast extraordinary (X) electromagnetic waves
are reflected at or below the critical region where $V=1$, except at a very narrow
angular range around the critical angle of incidence (the Spitze angle) 
where the O mode wave is mode converted to Z mode waves \citep{Budden61,Ginzburg67,Mjolhus1990,Gondarenko2003}.
(Throughout the manuscript, we use the standard notation $V=\omega_{pe}^2/\omega^2$, $Y=\omega_{ce}/\omega$
and $U=Y^2$ \citep{Budden61,Ginzburg67}
where $\omega_{pe}$ is the electron plasma frequency, $\omega_{ce}$
is the electron gyrofrequency and $\omega$
is the transmitted frequency.)
Enhanced topside plasma lines in the F2 region were first observed in
heating experiments at Arecibo by \citet{Ganguly1983}.
During experiments where the EISCAT UHF beam
was directed near the Spitze angle, \citet{Isham1990} observed enhancements
of ion lines in the topside interaction F2 region where
$V=1$. Similar observations at EISCAT were done in
the E region by \citet{Rietveld2002}, where there were no clear
dependence of the power of the topside echoes on the power of the
plasma turbulence at the bottomside.
There are several observations at EISCAT of topside echoes over an angular range
of several degrees \citep{Mishin2001}. In the experiment at EISCAT by \citet{Isham1999},
which was performed with low duty cycle (200 ms every 10 s),
topside excitation of ion and plasma lines were also observed in a angular
range of many degrees from the Spitze angle, including vertical incidence.
In order for normally incident O wave to tunnel through the F peak, conversion of
the O mode to the Z mode should occur, and the following conditions
are necessary. First, that the transmitter frequency
should be near the maximum plasma frequency of the F peak so that
$V<1+Y$ for the peak plasma frequency of the F layer, as discussed by
\citet{Mishing1997} and shown in the linear E-region simulations for
a parabolic density profile \citep{Gondarenko2003}. Second,
the existence of the density/electric field fluctuations
within the critical layer, which can occur due to the generation of
electrostatic turbulence excited by the parametric instability or in the earlier phase of
the nonlinear thermal self-focusing instability
\citep{Gondarenko2006}. \citet{Mishin2001} proposed that the topside 
plasma and ion lines could be due to resonant
scatter of the HF pump wave into Z mode waves on the small-scale
magnetic field aligned striations, which in turn are created by
the HF pump wave via the resonant instability, or, in the case of
the low duty cycle experiments with pump frequencies near a multiple
of the electron gyro frequency, due to a second-order four wave
interaction \citep{Huang1994}. Rocket observations during heating 
experiments in Arecibo \citep{Gelinas2003} revealed Z mode waves and 
field aligned striations above the O mode reflection point, and evidence
of generation of Bernstein waves where the local plasma frequency
matches an electron gyro harmonic.

During experiments where the EISCAT UHF beam was directed
along the magnetic field lines, \citet{Isham1996} discovered spectra 
with an unusually large spectral width (up to 200 kHz) and a mean 
frequency shifted by 200-300 kHz away from the heating frequency in 
both the upshifted and downshifted channels. This feature has been named 
the HF-induced outshifted line (HFOL). The HFOL spectra come from a height about 3-6 km above the
reflection height. \citet{Mishing1997} propose that the HFOL feature
can be consistently explained if the upper region is the reflection layer
of the Z mode appearing due to the mode conversion at the reflection
level of the injected O mode wave. Their scenario includes large-scale
density depletions at the Airy maxima of the Z mode, the parametric decay of 
the Z mode into electron Bernstein and lower hybrid waves \citep{Kuo1998}, 
and the production of suprathermal electrons by the electrostatic turbulence,
which in turn excite Langmuir waves inside the density depletions, 
yielding the HFOL.

The interaction between the electromagnetic pump wave and
the electrostatic turbulence have been modeled with Zakharov-type equations
\citep{Zakharov1972,Kuznetsov1974} and wave kinetic equations 
\citep{Fejer1973,Perkins1974}, to understand the wave
spectra and radiation generation from electromagnetically driven Langmuir turbulence 
\citep{Galeev1977,Mjolhus1995,DuBois1991,Stubbe1992}.
In this Letter we analyze the conversion of O mode waves
into Z mode waves and the generation of topside ionospheric
turbulence numerically with a full-scale
simulations of a generalized Zakharov model \citep{Eliasson2007} based on a
nested grid method \citep{Eliasson2007b}. We are using realistic length scales and ionospheric parameters
relevant for the high-latitude facility EISCAT
near Troms{\o}, Norway, but similar parameters may be also found at high-latitude facilities such as
HAARP, Alaska and HIPAS, near Fairbanks, Alaska.

\section{Mathematical model}

In our one-dimensional simulation geometry, we assume that a
large amplitude electromagnetic wave is injected vertically,
along the $z$ axis, into the vertically stratified ionospheric layer.
The mathematical model, derived in \citet{Eliasson2007}, is based on a
separation of time scales, where
the time-dependent quantities
(here denoted $\psi(z,t)$) are separated into a slow and a high-frequency
time scale, $\psi=\psi_s+\psi_h$. The high-frequency component
is assumed to take the form
$\psi_h=(1/2)\widetilde{\psi}(z,t)\exp(-i\omega_0 t)+$complex
conjugate, where $\widetilde{\psi}$
represents the slowly varying complex envelope of the high-frequency field,
and $\omega_0$ is the frequency of the transmitted electromagnetic wave.
Hence the time derivatives on the fast timescale will be transformed
as $\partial/\partial t\rightarrow \partial/\partial t - i\omega_0$ when
going from the $\psi_h$ variable to the $\widetilde{\psi}$ variable.
The dynamics of the high-frequency
electromagnetic waves and electron dynamics is coupled nonlinearly
to the slow timescale electron dynamics via the ponderomotive
force acting on the electrons, and is mediated to the ions via
the electrostatic field on the slow timescale.

We first present the envelope equations on the fast time scale.
The transverse (to the $z$ axis) components of the
electromagnetic field in the presence
of a slowly varying electron density $n_{es}$ is governed by the
electromagnetic wave equation
\begin{equation}
  \frac{\partial\widetilde{\bf A}_\perp}{\partial t}
  =i\omega_0\widetilde{\bf A}_\perp -\widetilde{\bf E}_\perp,
  \label{Eq_EM1}
\end{equation}
\begin{equation}
  \frac{\partial \widetilde{\bf E}_\perp}{\partial t}
  =i\omega_0\widetilde{\bf E}_\perp-c^2\frac{\partial^2\widetilde{\bf A}_\perp}{\partial z^2}+
  \frac{e n_{es} \widetilde{\bf v}_{{e}\perp}}{\varepsilon_0},
  \label{Eq_EM2}
\end{equation}
where $\widetilde{\bf A}_\perp$ and $\widetilde{\bf E}_\perp$
is the transverse vector potential (in Coulomb gauge) and electric field, respectively.
Here $c$ is the speed of light in vacuum and $\varepsilon_0$ is the electric vacuum permittivity.
The $z$ component of the electric field is governed by the Maxwell equation
\begin{equation}
  \frac{\partial \widetilde{E}_{z}}{\partial t}=i\omega_0\widetilde{E}_{z}+\frac{e n_{es}
  \widetilde{v}_{ez}}{\varepsilon_0}.
\end{equation}
The high-frequency electron dynamics is governed by the
continuity and momentum equation
\begin{equation}
 \frac{\partial \widetilde{n}_{e}}{\partial t}=i\omega_0
 \widetilde{n}_{e}-\frac{\partial(n_{es}\widetilde{v}_{ez})}{\partial z},
 \label{Eq_continuity2}
\end{equation}
and
\begin{equation}
\begin{split}
  \frac{\partial \widetilde{\bf v}_{e}}{\partial t}
  &=i\omega_0\widetilde{\bf v}_e-\frac{e}{m_{e}}\left(\widehat{\bf
  z}\widetilde{E}_z+\widetilde{\bf E}_\perp+\widetilde{\bf v}_{e}\times{\bf B}_0\right)
  \\
  &-\widehat{\bf z}\frac{3 v_{Te}^2}{n_0}\frac{\partial \widetilde{n}_{e}}{\partial z}
  -\nu_{e}\widetilde{\bf v}_{e},
\end{split}
  \label{Eq_momentum2}
\end{equation}
respectively, where ${\bf B}_0$ is the geomagnetic field,  
$\nu_{e}$ is the electron collision frequency, $e$ is the magnitude
of the electron charge, $m_e$ is the electron mass, $v_{Te}=(k_{B} T_e/m_e)^{1/2}$ is the
electron thermal speed, $k_B$ is Boltzmann's constant, $T_e$ is the electron temperature,
and $n_0$ is the background electron number density.
We have assumed that the ions are immobile on the fast time scale due to their large mass
compared to that of the electrons.

The slowly varying electron number density is separated as
$n_{es}=n_{e0}(z)+n_s(z,t)$ where $n_{e0}(z)=n_{i0}(z)$ represents the
large-scale electron density profile due to ionization/recombination
in the ionosphere, and $n_s$ is the slowly varying electron density
fluctuations. We regard the electrons as inertialess on
the slow time scale, where we also assume quasi-neutrality
$n_{is}=n_{es}\equiv n_s$. The dynamics of the ions is governed by
the continuity equation
\begin{equation}
  \frac{\partial n_s}{\partial t}+n_0\frac{\partial v_{iz}}{\partial
  z}=0,
  \label{Eq_IA_cont}
\end{equation}
and, since the ions are assumed to be unmagnetized, the slow timescale plasma velocity
is obtained from the ion momentum equation driven by the
ponderomotive force
\begin{equation}
  \frac{\partial v_{iz}}{\partial t}=-\frac{C_s^2}{n_0}\frac{\partial n_s}{\partial
  z}-\nu_i v_{iz}-\frac{\varepsilon_0}{4m_i n_0}\frac{\partial |\widetilde{\bf E}|^2}{\partial z}
  \label{Eq_IA_mom}
\end{equation}
where $C_s=[k_B(T_e+3T_i)/m_i]^{1/2}$ is the IA speed, $T_i$ is the ion
temperature, $m_i$ is the ion mass, and we have denoted
$|\widetilde{\bf E}|^2=|\widetilde{\bf E}_\perp|^2+|\widetilde{E}_z|^2$ and $|\widetilde{\bf
E}_\perp|^2=|\widetilde{E}_x|^2+|\widetilde{E}_y|^2$.
In obtaining Eq. (\ref{Eq_IA_mom}), we have used the ion momentum equation
\begin{equation}
\frac{\partial v_{iz}}{\partial t}=\frac{e}{m_{\bf i}}E_{zs}
- \frac{3v_{Ti}^2}{n_0}\frac{\partial n_{i}}{\partial z} -
\nu_{i}v_{iz},
\label{ion_mom}
\end{equation}
together
with the electron force balance equation
\begin{equation}
  0=-eE_{zs} -\frac{k_B T_e}{n_0}\frac{\partial n_s}{\partial z}-
  \frac{\varepsilon_0}{4n_0}
  \frac{\partial |\widetilde{\bf E}|^2}{\partial z},
\end{equation}
where we have neglected effects of the geomagnetic field ${\bf B}_0$ on
the ponderomotive force (the $|\widetilde{\bf E}|^2$-term)
since we will consider high-frequency waves with
frequencies much higher than the electron gyro-frequency.

\section{Numerical setup}

We next define the simulation setup and physical parameters
used in the simulation; the details of the numerical implementation
is described in the Appendix of \citet{Eliasson2007}.
The simulation code uses a one-dimensional geometry, along the $z$ axis.
Our simulation box starts at an
altitude of 200 km and ends at 400 km. While
$\widetilde{\bf A}_\perp$ and $\widetilde{\bf E}_\perp$ are represented on a grid with grid size
2 m everywhere, the rest of the quantities $\widetilde{n}_e$, $\widetilde{\bf v}_e$, $n_s$,
$v_{iz}$ and $\widetilde{E}_z$ are resolved with a much denser grid of grid size 4 cm
at the bottomside and topside interaction regions regions,
$z=286.2$--$287.2\,\mathrm{km}$ and $z=312.8$--$313.8\,\mathrm{km}$,
in order to resolve small-scale structures due to electrostatic turbulence.
Outside these regions, all quantities are resolved on the 2-meter grid.
This nested grid procedure is used to avoid a severe Courant-Friedrich-Levy (CFL) condition
$\Delta t < \delta z/c$ on the timestep if the electromagnetic field is resolved
on the dense grid $\delta z$. The ionospheric density profile will be assumed to
have a Gaussian shape of the form
$n_{i0}(z)=n_{0,max}\exp[-(z-z_{max})^2/10^9]$,
where $n_{0,max}=5\times10^{11}\,\mathrm{m}^{-3}$ and
$z_{max}=300\times10^3\,\mathrm{m}$ are the maximum density and the altitude of
the F peak, and $L=31.6\times10^3\,\mathrm{m}$ is the density scale length.
The external magnetic field $B_0$ is set to
$4.8\times 10^{-5}$ T and is tilted $\theta=0.2269$ rad ($13^\circ$)
to the vertical ($z$) axis so that
${\bf B}_0=B_0[\widehat{\bf x} \sin(\theta)-\widehat{\bf z} \cos(\theta)]$,
which is the case at EISCAT in Troms{\o}. We assume that the transmitter
on ground has been switched on at $t=0$. The simulation starts at $t=0.67\,\mathrm{ms}$,
when the electromagnetic wave has reached the altitude 200 km.
Initially, all time-dependent fields are set to zero, and random density
fluctuations of order $10^6\,\mathrm{m}^{-3}$ are added to $n_s$
to seed the parametric instability in the plasma. The electromagnetic
wave is injected from the bottomside of the simulation box at 200 km,
by setting the $x$ component of the electric field to 1 V/m on the
upward propagating field (see \citet{Eliasson2007}).
The transmitter frequency is set to $\omega_0=3.66\times10^7\,\mathrm{s}^{-1}$, which
is somewhat lower than the maximum F peak plasma frequency
$\omega_{pe,max}=3.99\times10^7\,\mathrm{s}^{-1}$, calculated with $n_0=n_{0,max}$.
On the other hand,
the chosen wave frequency is higher than the maximum Z mode cutoff
frequency $\omega_{Z,max}\sim3.58\times 10^7\,\mathrm{s}^{-1}$ calculated from $V=1+Y$
with $V=\omega^2_{pe,max}/\omega^2_{Z,max}$, $Y=\omega_{ce}/\omega_{Z,max}$,
and $\omega_{ce}=eB_0/m_e\approx8.5\times10^6\,\mathrm{s}^{-1}$.
By this choice of pump frequency, the O mode is reflected by the overdense plasma layer, while
Z mode waves are allowed to propagate to the topside when electrostatic turbulence is
generated due to the parametric instability.
We use oxygen ions so that $m_i=16\,m_p$ where $m_p=1836\,m_e$ is the proton mass.
Further, $T_e=T_i=1500\,\mathrm{K}$ so that $v_{Te}=1.5\times 10^5$ m/s and $C_s=1.75\times10^3$ m/s,
and we use $n_0=3.3\times 10^{11}\,\mathrm{m}^{-3}$, which is the
electron number density at $V=1$.
The electron collision frequency
is set to $\nu_e=10^3\,\mathrm{s}^{-1}$ and the effective ion ``collision frequency,''
due to Landau damping, is set to
$\nu_i=2\times 10^{3}\,\mathrm{s}^{-1}$, where we have neglected that the ion
Landau damping depends on the wavelength of the IA wave. The value
used here is approximately valid for IA waves with wavelengths of $\sim$1--2 m.
\section{Numerical results}

\begin{figure}
\centering
\includegraphics[width=8.5cm]{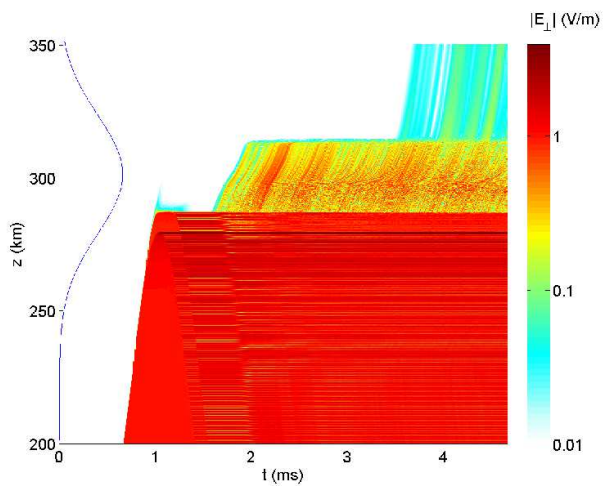}
\caption{The amplitude of the perpendicular (to the $z$ axis)
electric field in log scale and
the density profile of the F layer (solid line). Electrostatic
turbulence generated due to the parametric instability at the bottom
and top sides of the F layer: at $t\sim1.7\,\mathrm{ms}$, the
O wave is transferred into Z waves which tunnel through the
layer, and at $t\sim3.5$ ms, the inverse conversion process
yields O waves escaping the plasma.
}
\label{Fig_Eperp}
\end{figure}

\begin{figure*}
\centering
\includegraphics[width=15cm]{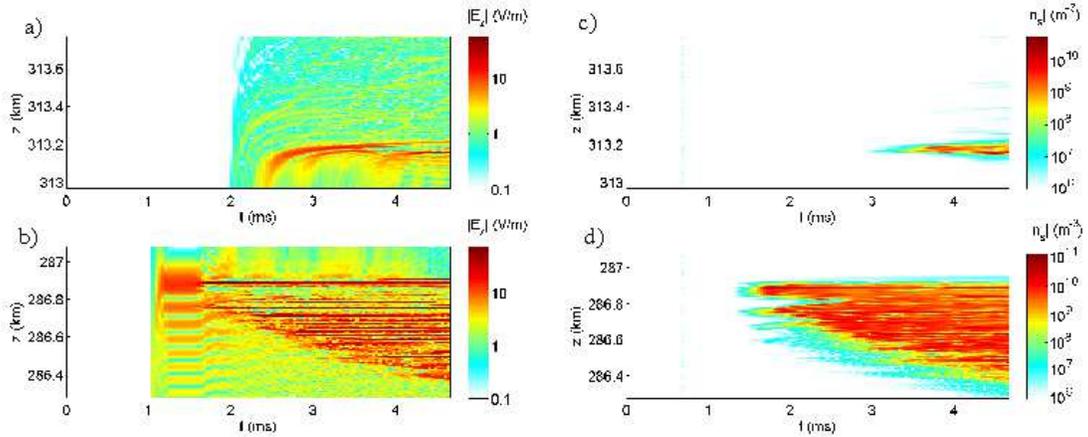}
\caption{
Electrostatic turbulence generated due to the parametric instability
shown with the electric field amplitude $E_z$ at (b) the bottomside and
(a) topside interaction regions of the F layer at $t\sim1.7$ ms and $t\sim3.5$
ms, respectively.
The ion density fluctuation $n_s$ at the interaction regions: (d) the
bottomside when large-scale fluctuations are excited due to wave
collapse and caviton formation at $t\sim1.7\,\mathrm{ms}$ and (c) the topside at
$t\sim3.5\,\mathrm{ms}$.
}
\label{Fig_Ezns}
\end{figure*}


The time evolution of the electric fields and ion density fluctuations are
shown in Figs. 1 and 2 from $t=0$ to $t=4.67\,\mathrm{ms}$.
Figure 1 shows the spatial and temporal evolution of the perpendicular
(to the $z$ axis) electric field
amplitude $|\widetilde{\bf E}_\perp|=\sqrt{|\widetilde{E}_x|^2+|\widetilde{E}_y|^2}$.
After approximately $1\,\mathrm{ms}$, the electromagnetic wave reaches the
bottomside interaction for the O mode at $z\approx 287\,\mathrm{km}$ and for the  X mode
at a slighter lower altitude $z\approx 270\,\mathrm{km}$,
after which a steady-state standing pattern quickly builds up. At $t\approx1.6\,\mathrm{ms}$,
there is an efficient conversion of the incident O wave into
the Z mode due to rapid development of the density fluctuations
(discussed in connection to Fig. 2 below). The Z mode waves propagate from the bottomside
at $z=270\,\mathrm{km}$ to the topside interaction region for the O
mode at $z=312\,\mathrm{km}$. At $t=3.5\,\mathrm{ms}$, we observe excitation
of O mode radiation at the topside at $z\approx 312\,\mathrm{km}$, and these O
mode waves propagate away from the ionospheric profile on the topside.

In Fig. 2, we have visualized the ${\widetilde E}_z$ component of the high frequency
electric field and the ion density fluctuations $n_s$, in the vicinity of the topside 
and bottomside interaction regions for the O mode. In Fig. 2(b), we see the arrival of 
the O mode at the bottom side at $t\approx 1\,\mathrm{ms}$, after which a standing wave pattern
 \citep{Ginzburg67} is set up. At $t\approx 1.7\,\mathrm{ms}$, the standing
wave pattern breaks up into small-scale, large amplitude waves. At the same
time, seen in Fig. 2(d), large amplitude ion density fluctuations are excited
first in the region of the maxima of $E_z$ at $z=286.75\mathrm{km}$ and 
$z=286.85\mathrm{km}$ (see Fig. 2), and later spreading to lower altitudes.
This transition into wave turbulence is closely correlated with the excitation of Z mode waves observed
in Fig. 1.  At $t=2$--$2.5\,\mathrm{ms}$, we see in Fig. 2(a) the arrival of the $Z$ mode at the
topside ionosphere. While some weaker waves propagate to higher
altitudes, a band of large-amplitude electric field is set up at $z\approx 313.2\,\mathrm{km}$
where the Z mode wave meets the transformation point \citep{Mjolhus1990,Mishin2001}
and converts into electrostatic waves. At $t$ later than $3.5\,\mathrm{ms}$ the
Z mode wave decays into short wavelength
Langmuir and IA waves due to a parametric instability, followed by Langmuir wave 
collapse and ion caviton formation. In Fig 2(c), the ion fluctuations are seen at
the topside in a narrow band at $313.2\,\mathrm{km}$.
The large amplitude collapsed Langmuir waves and the associated ion cavitons have sizes of a few
tens of centimeters or less and a spacing of a meter or less, similar as in the
study of \citet{Eliasson2007}. The topside Langmuir turbulence gives
rise to the radiation of O mode waves seen in Fig 1 at $t>3.5\,\mathrm{ms}$,  
in a process described by \citet{Mjolhus1995}.

\section{Discussion}

We have performed a simulation study of enhanced topside turbulence,
where the transmitter frequency is kept slightly
lower than the critical frequency of the F2 layer. Our simulation study reveals that
the electromagnetic wave decays parametrically on the bottomside of the ionosphere,
in which collapsing wavepackets and ion density depletions are created.
This is a three-wave decay of the O mode into one counter-propagating Langmuir (L)
wave and one forward scattered IA wave, obeying the matching conditions
$\omega_O=\omega_L+\omega_{IA}$ of the frequencies and $k_O=k_L+k_{IA}$ of the
wavenumbers, where $\omega_O(k_0)$, $\omega_L(k_L)$ and $\omega_{IA}(k_{IA})$
are given by the O mode, magnetized Langmuir, IA dispersion relation, respectively.
A wavelength of the O mode of the
order 100 meters leads to wavelengths of the Langmuir and IA waves
of the order one meter \citep{Eliasson2007}. The linear instability is followed
by a rapid Langmuir waves collapse and the generation of ion cavitons.
In the nonlinear phase of the instability, there is a strong generation 
of Z mode waves in the bottomside interaction region. These large-amplitude 
Z-mode waves are allowed to propagate into the denser plasma to the topside 
interaction region of the ionospheric layer, where they reach their 
transformation point and their amplitude increase as their group speed decreases and they are
turning into electrostatic waves. At the topside,
the Z mode waves undergo a parametric instability into small-scale 
(few tens of centimeters) coupled electrostatic and IA waves, 
followed by Langmuir wave collapse and the formation of ion cavitons.  
In this last step, Langmuir turbulence generated
O mode waves \citep{Mjolhus1995} are propagating away from the plasma layer on the top side.
We note that the wave tunneling and the excitation of topside turbulence occur
within a few milliseconds, and may explain the observed
topside enhanced plasma and ion lines far away from the Spitze region, in particular
during low duty-cycle ionospheric interaction experiments where the impact of
ionospheric irregularities are minimized \citep{Isham1999}.

\acknowledgments
This work was supported by the Swedish Research Council (VR).
Discussions with T. B. Leyser and B. Thid\'e are gratefully appreciated.



\begin{thebibliography}{}

\bibitem[{\it Budden}(1961)]{Budden61} Budden, K. G. (1961),
Radio waves in the ionosphere, Cambridge University Press.\vspace{0.2cm}



\bibitem[{\it Djuth et al.}(1990)]{Djuth1990} Djuth, F. T., M. P. Sulzer, and J. H. Elder (1990),
High resolution observations of HF-induced plasma waves in the ionosphere,
{\it Geophys. Res. Lett.}, {\it 17}(11), 1893--1896.\vspace{0.2cm}

\bibitem[{\it DuBois et al.}(1990)]{DuBois1990} DuBois, D. F., H. A. Rose, and D. Russell (1990),
Excitation of strong Langmuir turbulence in plasmas near critical density: Application to
HF heating of the ionosphere, {\it J. Geophys. Res.}, {\it 95}(A12), 21,221.\vspace{0.2cm}

\bibitem[{\it DuBois et al.}(1991)]{DuBois1991} DuBois, D. F., H. A. Rose, and D. Russell (1991),
Coexistence of parametric decay cascades and caviton collapse at subcritical densities,
{\it Phys. Rev. Lett.}, {\it 66}, 1970--1973.\vspace{0.2cm}

\bibitem[{\it Duncan and Sheerin}(1985)]{Duncan1985} Duncan, L. M., and J. P. Sheerin (1985),
High-resolution studies of the HF ionospheric modification region,
{\it J. Geophys. Res.}, {\it 90}(A9), 8371--8376.\vspace{0.2cm}

\bibitem[{\it Eliasson}(2007)]{Eliasson2007b} Eliasson, B. (2007), A nonuniform nested grid method
for simulations of RF induced ionospheric turbulence,
{\it Comput. Phys. Commun.}, {\it 178}, 8--14.\vspace{0.2cm}

\bibitem[{\it Eliasson and Stenflo}(2007)]{Eliasson2007} Eliasson, B., and L. Stenflo (2008),
Full-scale simulation study of the initial stage of ionospheric turbulence,
{\it J. Geophys. Res.}, {\it 113}, A02305, doi:10.1029/2007JA012837.\vspace{0.2cm}

\bibitem[{\it Fejer and Kuo},(1973)]{Fejer1973} Fejer, J. A., and Y.-Y. Kuo (1973),
Structure in the nonlinear saturation spectrum of parametric instabilities,
{\it Phys. Fluids}, {\it 16}, 1490, doi:10.1063/1.1694546.\vspace{0.2cm}

\bibitem[{\it Galeev et al.}(1977)]{Galeev1977} Galeev, A. A., R. Z. Sagdeev, V. D. Shapiro,
V. I. Shevchenko (1977) Langmuir turbulence and dissipation of high-frequency energy,
{\it Sov. Phys. JETP}, {\it 46}, 711--720.\vspace{0.2cm}

\bibitem[{\it Ganguly and Gordon}(1983)]{Ganguly1983} Ganguly, S., and W. E. Gordon (1983),
Heater enhanced topside plasma line,
{\it Geophys. Res. Lett.}, {\it 10}(10), 977--978.\vspace{0.2cm}

\bibitem[{\it Gelinas et al.}(2003)]{Gelinas2003} Gelinas, L. J., M. C. Kelley, M. P. Sulzer, E. Mishin,
M. J. Starks (2003),
In situ observations during an HF heating experiment at Arecibo: 
Evidence for Z-mode and electron cyclotron harmonic effects,
{\it J. Geophys. Res.}, {\it 108}(A10), 1382, doi:10.1029/2003JA009922.\vspace{0.2cm}

\bibitem[{\it Ginzburg}(1967/1970)]{Ginzburg67} Ginzburg, V. L.,
Rasprostranenie Elektromagnitnykh Voln v Plazme (The
Propagation of Electromagnetic Waves in Plasmas) 2nd ed. (Moscow:
Nauka, 1967) [Translated into English (Oxford: Pergamon Press, 1970)].\vspace{0.2cm}

\bibitem[{\it Gondarenko et al.}(2003)]{Gondarenko2003} Gondarenko, N. A., P. N. Guzdar, S. L. Ossakow,
and P. A. Bernhardt (2003), Linear mode conversion in inhomogeneous magnetized plasmas during ionospheric modification by HF radio waves, {\it J. Geophys. Res.}, {\it 108}(A12), 1470, doi:10.1029/2003JA009985.\vspace{0.2cm}

\bibitem[{\it Gondarenko et al.}(2006)]{Gondarenko2006} Gondarenko, N. A., S. L. Ossakow, and G. M. Milikh
(2006), Nonlinear evolution of thermal self-focusing instability in ionospheric
modifications at high latitudes: aspect angle dependence,
{\it Geophys. Res. Lett.}, {\it 33}, L16104, doi:10.1029/2006GL025916.\vspace{0.2cm}

\bibitem[{\it Huang and Kuo}(1994)]{Huang1994} Huang, J., and S. P. Kuo (1994),
Cyclotron harmonic effect on the thermal oscillating two-stream instability in the high latitude ionosphere,
{\it J. Geophys. Res.}, {\it 99}(A2), 2173--2181.\vspace{0.2cm}

\bibitem[{\it Isham et al.}(1990)]{Isham1990} Isham, B., W. Kofman, T. Hagfors,
J. Nordling, Bo Thid\'e, C. LaHoz, and P. Stubbe (1990),
New phenomena observed by EISCAT during an RF ionospheric modification experiment,
{\it Radio Sci.}, {\it 25}(3), 251--262.\vspace{0.2cm}

\bibitem[{\it Isham et al.}(1996)]{Isham1996} Isham, B., C. La Hoz, H. Kohl, T. Hagfors, T. B. Leyser, and M. T. Rietveld (1996),
    Recent EISCAT heating results using chirped ISR, {\it J. Atmos. Terr. Phys.}, {\it 58}, 369--383.\vspace{0.2cm}

\bibitem[{\it Isham et al.}(1999)]{Isham1999} Isham, B., B. T. Rietveld, T. Hagfors, C. La Hoz, E. Mishin,
W. Kofman, T. B. Leyser, and A. P. Van Eyken (1999), Aspect angle dependence of
HF enhanced incoherent backscatter, {\it Adv. Space Res.} {\it 24}(8), 1003--1006.\vspace{0.2cm}

\bibitem[{\it Kuo et al.}(1998)]{Kuo1998} Kuo, S. P., E. Koretzky, and M.C. Lee (1998),
Parametric excitation of lower hybrid waves by Z-mode waves near electron cyclotron harmonics at Troms{\o},
{\it J. Geophys. Res.}, {\it 103}(A10), 23,373--23,379.\vspace{0.2cm}

\bibitem[{\it Kuznetsov}(1974)]{Kuznetsov1974} Kuznetsov, E. A. (1974),
The collapse of electromagnetic waves in a plasma,
{\it Sov. Phys. JETP}, {\it 39}, 1003--1007. [{\it Zh. Eksp. Teor. Fiz.}, {\it 66}, 2037.]\vspace{0.2cm}


\bibitem[{\it Mishin et al.}(1997)]{Mishing1997} Mishin, E., T. Hagfors, and W. Kofman (1997),
On origin of outshifted plasma lines during HF modification experiments,
{\it J. Geophys. Res.}, {\it 102}(A12), 27,265--27,269.\vspace{0.2cm}

\bibitem[{\it Mishin et al.}(2001)]{Mishin2001} Mishin, E., T. Hagfors, and B. Isham (2001),
A generation mechanism for topside enhanced incoherent backscatter during high
frequency modification experiments in Troms{\o},
{\it Geophys. Res. Lett.}, {\it 28}(3), 479--482.\vspace{0.2cm}

\bibitem[{\it Mj{\o}lhus}(1990)]{Mjolhus1990} Mj{\o}lhus, E. (1990),
On linear conversion in a magnetized plasma,
{\it Radio Sci.}, {\it 20}, 1321--1339.\vspace{0.2cm}

\bibitem[{\it Mj{\o}lhus et al.}(1995)]{Mjolhus1995} M{\o}lhus, E., A. Hanssen, and D. F. DuBois (1995),
Radiation from Electromagnetically Driven Langmuir Turbulence,
{\it J. Geophys. Res.}, {\it 100}(A9), 17,527--17,541.\vspace{0.2cm}

\bibitem[{\it Perkins et al.}(1974)]{Perkins1974} Perkins, F. W., C. Oberman, and E. J. Valeo (1974),
Parametric instabilities and ionospheric modification, {\it J. Geophys. Res.}, {\it 79}(10), 1478--1496.\vspace{0.2cm}


\bibitem[{\it Rietveld et al.}(2002)]{Rietveld2002} Rietveld, M. T., B. Isham,
T. Grydeland, C. La Hoz, T. B. Leyser,
F. Honary, H. Ueda, M. Kosch, and T. Hagfors (2002), HF-pump-induced parametric instabilities in
the auroral E-region, {\it Adv. Space Res.}, {\it 29}(9), 1363--1368.\vspace{0.2cm}

\bibitem[{\it Stubbe et al.}(1992)]{Stubbe1992} Stubbe, P., H. Kohl, and M. T. Rietveld (1992),
Langmuir turbulence and ionospheric modifications,
{\it J. Geophys. Res.}, {\it 97}(A5), 6285--6297.\vspace{0.2cm}




\bibitem[{\it Zakharov}(1972)]{Zakharov1972} Zakharov, V. E. (1972),
Collapse of Langmuir waves, {\it Sov. Phys. JETP, Engl. Transl.}, {\it 35}, 908--912.

\end{thebibliography}
\end{document}